\documentclass[11pt]{article}

\usepackage[margin=1in]{geometry}
\usepackage[T1]{fontenc}
\usepackage{lmodern}
\usepackage{amsmath,amssymb}
\usepackage{booktabs}
\usepackage{graphicx}
\graphicspath{{./}{figs_bench/}{analysis/figures/}}
\usepackage{xcolor}
\usepackage{listings}
\usepackage[numbers,sort&compress]{natbib}
\usepackage[hidelinks]{hyperref}

\newcommand{\benchmark}{AxQM} % public name of the benchmark
\newcommand{\Ntasks}{1{,}019} % distinct holed statements
\newcommand{\Ntotalitems}{687} % numbered items extracted from N&C (NC/ROSTER.md)
\newcommand{\Nitems}{505} % items in the ledger
\newcommand{\Nnotasked}{26} % ledger items with no task (= \Nitems - \Nitemstasked)
\newcommand{\Nitemstasked}{479} % items carrying >= 1 task
\newcommand{\Nuniverse}{10{,}560} % deletable project-added declarations
\newcommand{\Nkept}{3{,}519} % kept set
\newcommand{\Nwithprereq}{387} % tasks with >= 1 prerequisite item (= \Ntasks - \Nnoprereq)
\definecolor{leankw}{RGB}{0,90,160}
\definecolor{leancomment}{RGB}{100,100,100}
\lstdefinelanguage{Lean}{
 morekeywords={theorem,lemma,def,structure,instance,variable,namespace,end,
 open,noncomputable,section,import,sorry,by,exact},
 sensitive=true,
 morecomment=[l]{--},
 morecomment=[s]{/-}{-/},
}
\newcommand{\ket}[1]{|#1\rangle}

\title{\benchmark: A Textbook-Scale Benchmark for Formal Proof Synthesis in a Library of Finite-Dimensional Quantum Mechanics}

\author{%
 Weichen Winston Yin$^{1}$, Jacob M. Taylor$^{1}$, Dirk R. Englund$^{1,3,*}$, Frank H.L. Koppens$^{2,4,*}$
 \\[6pt]
 \begin{minipage}{0.92\textwidth}\centering\small\itshape
 $^1$Axiomatic AI. $^2$Institut de Ci\`encies Fot\`oniques (ICFO). $^3$Massachusetts Institute of Technology (MIT). $^4$Instituci\'o Catalana de Recerca i Estudis Avan\c{c}ats (ICREA). $^{*}$Corresponding author(s): dirk@axiomatic-ai.com, frank.koppens@icfo.eu
 \end{minipage}%
}

\date{\today}

\begin{document}
\maketitle

\begin{abstract}
Formalizing mathematics in a proof assistant, where a machine checks every definition, statement and proof, has set a new standard of rigor. Large language models are now capable of formalizing autonomously, even at the scale of whole textbooks. We bring this standard of rigor to physics, where theoretical arguments carry idealizations that are rarely stated fully, and any logical gaps could have a cascading effect on interdependent results. Recognizing the need to evaluate autoformalization systems for physics, we release \benchmark, \Ntasks{} kernel-checkable proof-synthesis tasks over \Nitemstasked{} items drawn from the textbook \emph{Quantum Computation and Quantum Information} by Nielsen and Chuang. The tasks are stated in a custom Lean library of finite-dimensional quantum mechanics. By task count, it is the largest proof-synthesis benchmark in physics by a factor of four. \benchmark~is derived from a near-complete formalization of the formal portions of the textbook, so every task is guaranteed a solution, which we keep private. Grading of the benchmark is done deterministically by the Lean kernel, which checks that the proof compiles, that no \texttt{sorry} appears in it or in any declaration it depends on, and that it introduces no new axioms.
\end{abstract}

% ============================================================================
\section{Introduction}
\label{sec:intro}
Formalization in a proof assistant is increasingly used by the mathematics community to achieve machine-checked rigor that eliminates the possibility of an incomplete proof. Lean~4 \citep{moura2021lean4} is one such assistant: a programming language in which definitions, theorem statements and proofs are all written in code, and whose compiler accepts a proof only when a small trusted kernel has checked every step of it. Faced with the proof of a theorem formalized in Lean, a reviewer can now simply check that the statement carries the correct mathematical meaning, and fully trust that the proof of that statement has been checked down to the axioms by the kernel. The whole burden of review becomes a single question rather than pages upon pages of argument: does this formal statement say what it claims to say? With autoformalization, the translation of mathematical text into Lean code can now be performed by AI at scale, and what is produced this way has grown from single theorems to entire textbooks \citep{gloeckle2026textbook,wang2026m2f,rammal2026formalizing}. LLM formal provers for Lean now span reinforcement-learning systems \citep{xin2024deepseekprover,wang2025kimina,chen2025seedprover,achim2025aristotle,lin2025goedelv2,hubert2025alphaproof}, retrieval-augmented assistants \citep{yang2023leandojo,song2024leancopilot}, and agentic scaffolds around general-purpose models, including one tool-assisted iterative prover \citep{breen2025axprover} and a minimal open-source baseline \citep{requena2026minimal}.

Outside of mathematics, physics stands to gain a great deal from this technological advancement. Formal reasoning in the physical sciences, machine-checked and traceable to physical foundations, is now within reach. In some ways, physics needs this more than mathematics does. Seldom is physical reasoning expressed at the level of mathematical rigor \citep{toobysmith2026perspective}. Everything from quantum field theory to adiabatic elimination is full of assumptions and occasional hand-waving, yet remains sufficient for predicting the outcomes of experiment. Where the reasoning is formalizable, formalization has yielded fruitful results: formalizing the stability conditions of the two-Higgs-doublet potential recently exposed an error in a widely cited paper \citep{toobysmith2026higgs}, and a machine-verified proof has settled an open conjecture in quantum optimization \citep{kol2026quantum}. By moving to AI-assisted autonomous formalization, physicists are confronted up front with new challenges, but can also benefit from advances in \emph{verifiable} computer-aided reasoning.

Crucially, formalization must build upon a well-scaffolded set of mathematical axioms and definitions that self-consistently enable lemmas and theorems that build upon that foundation. To date, progress in this scaffolding is largely available only in a few branches of mathematics through the community effort known as Mathlib \cite{mathlib2020}. Physics has begun to acquire the same scaffolding.

PhysLean, formerly HepLean, collects formalized results across classical mechanics, relativity, condensed matter, quantum field theory and string theory \citep{toobysmith2024heplean,toobysmith2026perspective}, with supporting developments in index notation and Wick's theorem \citep{toobysmith2024index,toobysmith2025wick} and an axiomatization of quantum field theory \citep{douglas2026qft}. Formalization is being attempted across physics more broadly, from chemical physics \citep{bobbin2022chemical} and the mean-field derivation of the Vlasov equation \citep{miller2026vlasov} to control theory \citep{doll2026control}, numerical analysis \citep{tekriwal2021lax}, power-flow analysis \citep{khanpour2026powerflow} and particle-physics model building \citep{krippendorf2026physics}, with related work in other proof assistants \citep{huerta2024scalable}. In quantum information, Lean-QIT builds an operational layer for quantum Shannon theory, while Lean-Quantum builds it in a basis-independent way; Lean-QuantumInfo covers finite-dimensional quantum information in over a thousand theorems \citep{zhu2026leanqit,kasaura2026leanquantum,meiburg2025quantuminfo,meiburg2025stein}.

Individual formalized results in quantum information include Shor's algorithm \citep{peng2022shor,zhang2026shor}, the generalized quantum Stein's lemma \citep{meiburg2025stein}, CHSH rigidity \citep{zhao2026chsh}, error correction \citep{ehatamm2026qec}, and the fundamental theorem of matrix-product states of tensor network theory \citep{lu2026tensor}, several of them assisted by AI. Agentic autoformalization systems have also been built for the domain, targeting certified quantum neural network design and quantum computation broadly \citep{jing2026qnn,ren2026merlean}. Quantum computation and protocols have also been formalized in other proof assistants: in Coq \citep{boender2015quantum,paykin2017qwire,hietala2021voqc,zhou2023coqq} and in Isabelle/HOL \citep{bordg2020certified,bordg2020dirac,echenim2023chsh}.

Here we go beyond individual formalized results and apply autonomous formalization to a physics textbook in full, releasing \benchmark, a Lean~4 benchmark for the formalization of quantum information and computing theory, built from the exercises in the textbook \emph{Quantum Computation and Quantum Information} by Nielsen and Chuang \citep{nielsen2010quantum}. Following the scope of the textbook itself, this benchmark is purposely limited to finite-dimensional Hilbert spaces. Every construction therefore stays inside Mathlib's finite-dimensional linear algebra library, where the spectral theorem, the trace, and tensor products are available without functional-analytic hypotheses. Infinite-dimensional systems and unbounded operators are outside this scope.

\begin{figure}[t]
\centering
\includegraphics{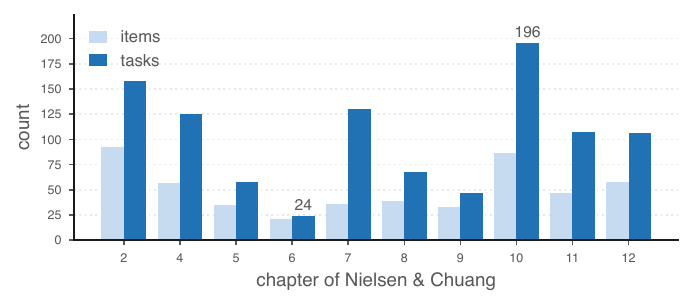}
\caption{Coverage: items and tasks by chapter of Nielsen and Chuang. Items refer to explicitly numbered items in the book: theorems, problems, exercises, and examples. Tasks are individual Lean theorem statements that capture all or part of an item. Each item may have multiple tasks. Chapters 1 (overview) and 3 (classical computation) contribute no items to the benchmark. Chapters 7 (physical realization) and 10 (error correction) are the densest in tasks per item.}
\label{fig:coverage}
\end{figure}

\section{The Benchmark}
\label{sec:benchmark}

\benchmark~consists of \Ntasks{} proof-synthesis \emph{tasks} over \Nitemstasked{} \emph{items}, covering a large majority of the textbook's exercises and theorems (Fig.~\ref{fig:coverage}). Each task is a single formal statement in Lean whose proof is to be filled in, and each item (\emph{e.g.~}Theorem~11.8, Exercise~12.15) can include one or more tasks that are intended to capture the item's full meaning. Topics include the density-operator formalism and the Schmidt decomposition, universal gate sets and circuit decompositions, the quantum Fourier transform with order-finding and Grover search, quantum channels and the distance measures on them, stabilizer codes and fault tolerance, and von Neumann entropy and its inequalities.

The tasks span a wide range of difficulty. At one end are one-line identities about a Pauli matrix; at the other are items, often given as exercises to students, whose formalization involves proofs not only outside the text but also outside Mathlib. For example, Exercise~9.9 asks for a fixed point of every quantum channel and points the reader to Schauder's theorem. Neither Schauder's nor Brouwer's theorem is in Mathlib, so the reference proof instead uses the linearity of the channel and a mean-ergodic averaging argument on the compact convex set of density operators. The task with the largest reference proof (by declaration count) introduced more than 450 declarations beyond the benchmark library. We assign each task a proof length estimate from this count, the number of declarations its reference proof needs beyond the benchmark library, on a five-point scale from ``very small'' to ``very large'' (Table~\ref{tab:length}).\footnote{This scale measures something different from the difficulty scale of \citet{zhang2026qitbench}. Of the five items Lean-QIT-Bench shares with \benchmark, four of them carry that benchmark's maximum difficulty of 10. On our scale, the same four fall into ``moderate'' for Exercise~9.9, ``large'' for Theorem~9.2 and Theorem~12.9, and ``very large'' for Theorem~12.1. \citet{zhang2026qitbench} caution that their 1--10 assignments are not objective measurements of proof complexity. Neither does our scale measure difficulty for a human or LLM to solve the problem.}

\begin{table}[t]
\centering
\begin{tabular}{@{}lrr@{}}
\toprule
Proof length estimate & Tasks & Share \\
\midrule
very small & 158 & 15.5\% \\
small      & 324 & 31.8\% \\
moderate   & 280 & 27.5\% \\
large      & 190 & 18.6\% \\
very large &  67 &  6.6\% \\
\midrule
Total      & \Ntasks{} & 100\% \\
\bottomrule
\end{tabular}
\caption{Proof length estimates for the \Ntasks{} benchmark tasks. The band is assigned from the number of declarations a task's reference proof needs beyond the released benchmark library. It
estimates the length of a reference proof, not how hard a task is for a human or a model to solve.}
\label{tab:length}
\end{table}

Existing benchmarks for formal proof synthesis (generating the proof code for a formal statement) are drawn primarily from competition mathematics, ranging in size from 488 to 5,560 statements \citep{zheng2022minif2f,tsoukalas2024putnambench,yu2025formalmath}. With textbooks as the source material, ProofNet samples 371 statements across many undergraduate books \citep{azerbayev2023proofnet}, and TaoBench contains 150 from one analysis textbook by Terence Tao \citep{taylor2026taobench}. Formal Conjectures collects formalized but unproved statements and grows continuously \citep{firsching2026formal}. SorryDB likewise maintains a live database of unproved \texttt{sorry} goals harvested from 78 public Lean projects \citep{letson2026sorrydb}, and VeriSoftBench builds repository-scale software-verification tasks \citep{xin2026verisoftbench}. Formal benchmarks in physics are still in a nascent stage, at 76 to 250 tasks \citep{zhang2026qitbench,breen2025axprover,li2025lean4physics,zhang2026physprover}. Formal proof translation has also been benchmarked across proof assistants \citep{wu2026itpeval}; the physics benchmarks that ask only for informal reasoning are larger but do not use the kernel checker \citep{chung2025tpbench,qiu2025phybench,xu2025ugphysics}. Our benchmark, \benchmark, differs from these in two ways. It is a census of a single book rather than a sparse sample drawn across many sources, and its tasks are presented as part of a Lean library instead of standing alone. Solving the tasks thus requires working within and building on a custom library of quantum physics outside of an LLM's training corpus.

\benchmark~includes the formal statements of items in the book expressed in a shared foundational library of quantum physics, on top of a forked version of Mathlib (justified in \S\ref{sec:forking}). The fork was needed because Mathlib's implementation of multilinear maps is not general enough to express the inner product on systems composed of several registers. A separate ``solution library'', which we keep private, holds the formal proofs of all benchmark tasks. We take these steps to ensure that every benchmark task is solvable, each task has a difficulty estimate, and the formal solutions are not leaked for LLM training \citep{jacovi2023stop,balloccu2024leak}.

Grading follows the usual standard of Lean benchmarks: the library compiles, the submitted proof does not have \texttt{sorry} in its dependencies, and no new axioms are added. As an indication of scale, the benchmark library adds \Nkept{} Lean declarations on top of the Mathlib fork, while the complete library with solutions adds \Nuniverse{}. The released benchmark keeps only the minimal set of declarations needed for the tasks to compile, without any of the infrastructure needed to prove them.

\begin{figure}[t]
\centering
\includegraphics{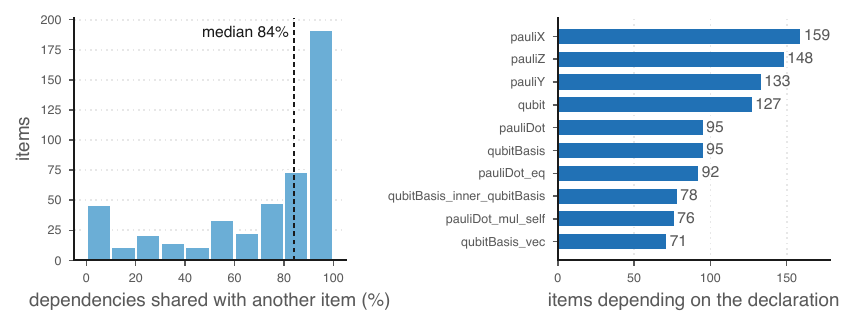}
\caption{Shared use of the quantum physics library underlying the benchmark, measured on the full solution library. \textbf{Left:} for each item, the share of its dependency closure (declaration count) that at least one other item also depends on; the median item shares 84\% of its dependencies with another item. \textbf{Right:} the ten most depended-upon declarations, by the number of distinct items whose statements reach them. A definitional error in any of these would have had to survive every one of those items' proofs.}
\label{fig:reuse}
\end{figure}

To further establish confidence in the quantum physics library underlying \benchmark, we develop metrics to clarify how much of the library is reused by multiple items. For example, one could ask what percentage of the Lean declarations in the dependency closure of a formalized textbook item are also used by at least one other item. For the median item in the solution library, this number is 84\% (Fig.~\ref{fig:reuse}). Foundational concepts are also highly reused throughout the library: each of the Pauli matrices is used by well over a hundred items, as many as 159. While this in no way guarantees the semantic correctness of the library, the high interdependency between declarations and the successful formalization of a large number of textbook items across a range of topics constrain the kinds of errors that can still persist.

This is also how conventional science is built. A physicist does not re-derive the spectral theorem or the no-cloning theorem before using them. A result is established once and then relied on, and its reliability grows with every independent use of it in downstream results, a process that leaves its trace in the citation record. Our library demonstrates that process explicitly in the formal relationships between Lean declarations. The declarations with the most dependencies were introduced early and gathered dependents monotonically without being restated. Critically, our approach is well designed for textbooks, where this ordering of dependencies is a typical pedagogical choice.

\begin{figure}[t]
\centering
\includegraphics{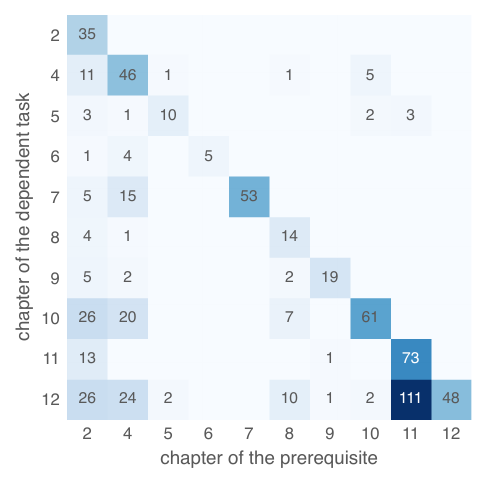}
\caption{Direct dependencies between tasks, aggregated by chapter, measured on the full solution library. For example, there are 15 instances of tasks in chapter 7 directly depending on tasks in chapter 4. The matrix is close to lower-triangular, which reflects the pedagogical structure of the book, where later chapters build upon earlier chapters.}
\label{fig:dag}
\end{figure}

This interdependency is also seen between individual items. In the solution library, \Nwithprereq{} of the \Ntasks{} tasks were proved by invoking at least one other task. This suggests that the benchmark can be run under two regimes: the \textbf{independent regime} requires each task to be completed on its own, and the \textbf{dependency-order regime} requires the prerequisite tasks to be completed first. We record the empirical proof dependencies between tasks in the published ledger; the aggregate dependencies between chapters of the book are given in Fig.~\ref{fig:dag}.

At \Ntasks{} tasks, \benchmark~is the largest benchmark for formal proof synthesis in physics to date. Purpose-built proof-synthesis benchmarks in physics have not exceeded 200 tasks \citep{li2025lean4physics,breen2025axprover,zhang2026qitbench}, while the largest comparable evaluation set is the 250-instance held-out split of PhysLeanData, a training corpus harvested from PhysLean \citep{zhang2026physprover}. This puts \benchmark~at 4.1 times the size of PhysLeanData. Even if one counts individual textbook items, each formalized as one or more tasks, there are still \Nitemstasked{}. We expect that \benchmark, as a formalization of a significant portion of a physics textbook, sets a baseline for the formalization of a field of physics.

It is worth contrasting our proof-synthesis benchmark with benchmarks of a different class: a model is asked to produce a faithful formal \emph{statement} rather than a proof of one. FormalPhysics has 200 problems, graded by compilation and an LLM judge \citep{meadows2026formalscience}, while QuantumLean-Bench, which has 931 problems, is only graded by a manual faithfulness rubric that awards credit even when the Lean code does not compile \citep{goswami2026quantumlean}.

While we are confident in the quality of the benchmark items, we welcome the community's thorough review of their physical correctness and faithfulness to the book. This is the failure mode that matters for a formal benchmark, and recent audits of widely used Lean suites show it is not rare: a statement can be vacuous, or quietly weaker than the claim in print, and still compile and still be provable \citep{ammanamanchi2026faults}. The Lean kernel is silent on such issues by construction. We give the structural argument that constrains such errors above (Fig.~\ref{fig:reuse}), and record what it does not cover in \S\ref{sec:failure}. Any corrections and improvements will be part of periodic updates to the benchmark following the initial release. Feedback may be sent by contacting the authors directly or by opening issues on the benchmark's GitHub repository.

%%%%%%
\section{Forking Mathlib}
\label{sec:forking}
Many proof-synthesis benchmarks take a pinned version of Mathlib as a package dependency \citep{tsoukalas2024putnambench,yu2025formalmath,hu2024minictx,zhang2026qitbench,taylor2026taobench}. This allows tasks to be stated and attempted with the full mathematical machinery provided by Mathlib. As a community-curated library with a stringent review process, Mathlib is usually taken as a trusted layer in formalization projects. We should therefore justify our choice to base our quantum physics library on an \emph{edited} version that forked off \texttt{leanprover-community/mathlib4} at commit \texttt{e560e3ad}, on Lean toolchain \texttt{v4.30.0-rc1}, and explain the content of the edits.

All 24 file changes we made to Mathlib are part of a single refactor. In the official Mathlib, \texttt{MultilinearMap} is linear over a single ring in each of its arguments. Our fork generalizes it to a multi-\emph{semi}-linear map, so that scaling one argument by $c$ scales the value by $\sigma(c)$, where $\sigma$ is a ring homomorphism. The original Mathlib statement is recovered by setting $\sigma$ to be the identity homomorphism.

The generalization is forced by the foundations of finite-dimensional quantum mechanics themselves. It is needed to define an inner product on the $n$-ary tensor product of quantum state spaces, an inner product that is conjugate-linear in each argument of the left factor ($\sigma$ being complex conjugation). A number of benchmark items concern multi-party registers, which are such tensor powers, and this refactor enables us to apply Mathlib's trusted \texttt{InnerProductSpace} machinery to them, without developing a new API for multi-conjugate-linear maps separate from \texttt{MultilinearMap}.

This refactor closely follows a precedent Mathlib has already set. \texttt{LinearMap} was generalized to a semilinear map along an arbitrary ring homomorphism \citep{dupuis2022semilinear}, and the inner product on a \emph{binary} tensor product was then stated in terms of it. Our change is the $n$-ary analog of that construction. As of the writing of this text, the same refactor is an open pull request under review on Mathlib (\#42534). If it is accepted, our benchmark can then be rebased to a newer pinned version of Mathlib in a future update, removing the need for the fork.
%%%%%%

\section{Failure modes in formalizing physics}
\label{sec:failure}
%%%%%%
In this section, we describe a few failure modes in the formalization of physics and clarify what kinds of errors can evade a proof assistant's kernel. Some of these are shared by the formalization of mathematics \citep{meek2026formalizing,mohammad2026faithfulness,klowden2026}. The most elementary one is that the formal statement does not mean what it purports to mean. The statement ``one plus one is equal to two'' can be formalized as \texttt{One + One = Two}, with both \texttt{One} and \texttt{Two} incorrectly defined as the natural number 0. The statement is nevertheless provable in Lean, but it means something entirely different.

There is a more subtle manifestation of this error in the formalization of natural language text that may be more pronounced in physics. We first illustrate it with an elementary example. Suppose the natural language source text reads, ``show that a particle undergoing constant acceleration starting from rest travels a distance that is quadratic in time.'' The following Lean code purports to formalize and prove this statement:

\begin{lstlisting}
def distanceTraveled (a t: (*$\mathbb{R}$*)): (*$\mathbb{R}$*):= a * t ^ 2 / 2

theorem distanceTraveled_eq (a t: (*$\mathbb{R}$*)) :
    distanceTraveled a t = a * t ^ 2 / 2 := rfl
\end{lstlisting}

This compiles, contains no \texttt{sorry}, and adds no new axiom. From the standpoint of the Lean kernel, this is a ``correct'' formalization. However, its physical content is vacuous: it merely defines the ``distance traveled'' to be $\frac{1}{2}at^2$, and proves that it is equal to $\frac{1}{2}at^2$ by definition. Nevertheless, in isolation, each declaration can be seen as semantically correct.

In contrast, a faithful formalization of the natural language statement would have to introduce the trajectory as a formal object constrained by the physical hypotheses: a function $x: \mathbb{R} \to \mathbb{R}$ whose second derivative is the constant $a$, with $x(0) = 0$ and $x'(0) = 0$, and then show that $x(t) = \frac{1}{2}at^2$. The difference between this and the vacuous version is the entire content of the exercise, the integration of the equation of motion, but the gap is invisible to the Lean kernel.

Physics is more exposed to this than mathematics for a structural reason. Objects in a mathematical text are already mathematical objects, usually with a clear definitional chain back to a set of canonical and rigorous definitions. Physics texts, however, speak of both the physical objects and the mathematical objects that model them, often in an interchangeable way. This modeling step is not something a proof assistant can check, even in principle, and is also one that sometimes confuses the typical student learning physics.

While the above error states the conclusion as a definition, a related error moves the conclusion, or a critical logical step, into the hypotheses of a theorem, which weakens the statement and avoids proving that step. These kinds of errors can be found in existing proof-synthesis benchmarks for quantum information theory, Lean-QIT-Bench and Lean-QuantumAlg-Bench, which publish 76 tasks \citep{zhang2026qitbench}.\footnote{\texttt{QudeLeap/Lean-QuantumAlg-Bench} at \texttt{b540e86} and \texttt{QuAIR/Lean-QIT-Bench} at \texttt{231830d}, both 2026-08-15.} We list two of these errors.

In their task \texttt{HamiltonianSimulation/FirstOrderLieTrotterGlobalErrorScaling}, the informal statement asks for a proof that a function $f(m)$ asymptotically scales as $O(1/m)$. However, the formal statement in the benchmark is (schematically) ``$\cdots \forall m \cdots \exists K \cdots f(m) \leq K/m$''. The ordering of the logical quantifiers means that $K$ is allowed to depend on $m$, making the statement trivially true by setting $K = m f(m)$. This is certainly not the intention of the informal statement, but this error is not caught by the Lean kernel checker.

Another task, \texttt{Fourier/QPESuperpositionExactEigenvectors}, asks for a proof that phase estimation, run on the superposition $\alpha\ket{u_1}+\beta\ket{u_2}$ of two eigenvectors, produces $\alpha\ket{N\phi_1}\ket{u_1}+\beta\ket{N\phi_2}\ket{u_2}$. However, the phase estimation circuit is never constructed or referred to in the formal statement. Instead, the circuit's action on an arbitrary eigenvector is taken as a hypothesis, bypassing the algorithm entirely. The task is therefore solved trivially by linearity. This semantic deficiency again evades the Lean kernel.

We have taken care to minimize these types of defects in \benchmark, but a thorough evaluation can only be done by relying on a larger effort beyond the authors. We invite the community of physicists and Lean experts to inspect and review the content of \benchmark, with any resulting improvements included in future updates to the benchmark.

\section{Discussion}
\label{sec:discussion}

While physics heavily invokes and relies on mathematical reasoning, it is reasonable to ask to what extent physical reasoning as a whole could be formalized. \citet{douglas2026axioms} points out that in \emph{mathematical physics}, a rigorous statement and proof already exist, and thus their formalization may proceed exactly as in mathematics proper. Much of theoretical physics, however, is not of this kind, with results often justified by arguments that a community trusts without a rigorous statement or proof (such as perturbative quantum field theory via path integrals). Douglas's response is to extract the rigorous parts of the physical reasoning as explicit mathematical premises and consequences, so that physics enters only through typed hypotheses and a dictionary between physical quantities and mathematical objects.

An example of this split is conventional (BCS) superconductivity. The theory assumes an effective attraction between electrons and restricts the many-body problem to mean-field states. From the resulting BCS functional onward, all further steps can be derived with mathematical rigor: the gap equation, the energy gap and the transition temperature~\citep{hainzl2008bcs,frank2012ginzburg}. The first two steps are physical assumptions: the attraction is an
effective interaction obtained from a well-established but not mathematically rigorous calculation, and the mean-field restriction is an approximation which is only partially justified~\citep{braunlich2014quasifree}. In Douglas' terms, those two steps are the typed hypotheses, and the identification of the measured gap with the order parameter of the functional is a dictionary entry.

\benchmark~applies the same split to a physics textbook. The postulates of quantum mechanics enter the library as its primitives, and each formal statement is a dictionary entry that ties a claim in the textbook to a mathematical object. The kernel checks every proof from there on. The dictionary is also where an error can enter. \citet{toobysmith2026perspective} states that an interactive theorem prover verifies logical consistency but cannot prevent a physically meaningless starting point, so whether each formal statement is faithful to the textbook rests on an expert's review, and the kernel does not check it (\S\ref{sec:failure}).

The first decision is therefore which parts of the textbook can be dictionary entries at all. Those are entries that are expressed, or expressible, as rigorous mathematical statements. Out of the \Ntotalitems{} items extracted from the textbook, \Nitems{} are formalized; the other items (essay prompts, graph plotting, intuitive arguments, numerical computation, research problems, and infinite-dimensional quantum mechanics statements) are excluded. Out of the formalized items, \Nitemstasked{} have tasks in this benchmark. The remaining \Nnotasked{} items were excluded for one of five reasons: the item is an existing theorem in upstream Mathlib, its proof is needed for the statement of another task to compile, it is a definition with no proof to synthesize, it duplicates another item, or we removed it manually for quality. \benchmark~and its solution library are therefore not a complete formalization of every physically meaningful statement in the textbook. Such a formalization is out of reach in principle, since the items excluded first are not mathematical statements, and we narrowed it further in practice by the five conditions above. \benchmark~is nonetheless the first textbook-scale formalization of physics (\S\ref{sec:benchmark} compares its size with other proof-synthesis benchmarks in physics).

We intend for \benchmark~to serve as a baseline for the evaluation of proof synthesis in physics. We chose \citet{nielsen2010quantum} as the source text because of the textbook's foundational status in the field of quantum information and quantum computing. As a pedagogical book, it naturally includes many problems which could be considered ``easy'' for a state-of-the-art LLM. We also include these problems in the benchmark, so that the benchmark score is an accurate measure of the model's ability to complete a course based on this textbook.

%%%%%%

\section{Data and code availability}
\label{sec:availability}

\benchmark~is published as a GitHub repository at \url{https://github.com/Axiomatic-AI/AxQM} under the Apache 2.0 license. The repository contains the task statements, the finite-dimensional quantum mechanics library, the Mathlib fork (branched from \texttt{leanprover-community/mathlib4} at commit \texttt{e560e3ad}, Lean toolchain \texttt{v4.30.0-rc1}), the per-task proof length estimates, the task-dependency ledger, and the grading and submission scripts. The solution library, containing reference proofs of all \Ntasks{} tasks, is withheld to keep the solutions out of model training corpora. An overview of the benchmark structure and instructions for running and grading it are included in the repository.

Comments and error reports may be submitted as issues on the GitHub repository. Inquiries and benchmark submissions may be sent to the authors by email.

% ============================================================================
\section*{Acknowledgments}
We thank Luigi Massacci, Michail Karatarakis and Victor Galitski for their critical inspection of the code and for checking the faithfulness of formal statements to the textbook.

\section*{Author contributions}
 W.W.Y.\ produced the foundational library of quantum physics on which the benchmark is built, contributed to the development and testing of the agentic autoformalization pipeline which aided in the construction of the benchmark, performed the quality check and preparations for the benchmark's release, and produced the figures. J.M.T.\ reviewed the benchmark code and checked formal statements against the textbook for faithfulness. D.R.E.\ proposed and carried out initial work on agentic formalization grounded in the axioms of quantum mechanics, and its specialization to quantum information and to Nielsen and Chuang in particular, including the restriction of the library to finite-dimensional Hilbert spaces. F.H.L.K.\ built the first agentic harness for autonomous formalization of the Nielsen and Chuang textbook, produced the prototype of the foundational library of quantum physics grounded in the axioms of quantum mechanics, and, with W.W.Y., developed the harness that built the library at scale. All authors discussed the results and contributed to the manuscript.

\section*{Competing interests}
W.W.Y., D.R.E. and J.M.T. are affiliated with Axiomatic AI, and F.H.L.K. is a co-founder of Axiomatic AI.

% ============================================================================

% ============================================================================
\bibliographystyle{unsrtnat}
\bibliography{refs/references}

\end{document}